\input harvmac
\input amssym
\baselineskip 13 pt

\def\p{\partial}

\def\rt{\rightarrow}

\def\zb{\overline{z}}

\def\psib{\overline{\psi}}

\def\taub{\overline{\tau}}

\def\eps{\epsilon}

\def\Ab{\overline{A}}
\def\Fb{\overline{F}}

\def\zb{\overline{z}}
\def\Lc{{\cal L}}
\def\Lcb{\overline{\cal L}}

\def\Wc{{\cal W}}
\def\Wcb{{\overline{\cal W}}}

\def\alphab{\overline{\alpha}}

\def\phib{\overline{\phi}}
\def\alphab{\overline{\alpha}}
\def\betab{\overline{\beta}}
\def\barn{\overline{n}}
\def\betab{\overline{\beta}}
\def\barb{\overline{b}}

\def\rar{\rightarrow}

\def\p{\partial}
\def\Wc{{\cal{W}}}
\def\half{{1\over 2}}
\def\rar{\rightarrow}
\def\mw{{\cal{W}}}
\def\ml{{\cal{L}}}

\def\hsl{hs[$\lambda$]}
\def\sln{SL(N,R)}
\def\slt{SL(2,R)}
\def\slth{SL(3,R)}

\def\vs{\vskip .1 in}
\def\a{\alpha}
\def\o{\omega}
\def\l{\lambda}
\def\t{\tau}

\def\mwil{${\mw}_{\infty}[\l]$}
\def\mwio{${\mw}_{\infty}[1]$}
\def\cg{{\cal{G}}}
\def\paps{\refs{\GutperleKF,\AmmonNK}}
\def\papa{\GutperleKF}
\def\papb{\AmmonNK}

\lref\CampoleoniZQ{
  A.~Campoleoni, S.~Fredenhagen, S.~Pfenninger and S.~Theisen,
  ``Asymptotic symmetries of three-dimensional gravity coupled to higher-spin
  fields,''
  JHEP {\bf 1011}, 007 (2010)
  [arXiv:1008.4744 [hep-th]].
}

\lref\BanadosGG{
  M.~Banados,
  ``Three-dimensional quantum geometry and black holes,''
  arXiv:hep-th/9901148.
}

\lref\BanadosNR{
  M.~Banados and R.~Caro,
  ``Holographic Ward identities: Examples from 2+1 gravity,''
  JHEP {\bf 0412}, 036 (2004)
  [arXiv:hep-th/0411060].
}

\lref\KrausVU{
  P.~Kraus, F.~Larsen and A.~Shah,
  ``Fundamental Strings, Holography, and Nonlinear Superconformal Algebras,''
  JHEP {\bf 0711}, 028 (2007)
  [arXiv:0708.1001 [hep-th]].
}

\lref\KrausNB{
  P.~Kraus and F.~Larsen,
  ``Partition functions and elliptic genera from supergravity,''
  JHEP {\bf 0701}, 002 (2007)
  [arXiv:hep-th/0607138].
}

\lref\DHokerHR{
  E.~D'Hoker, P.~Kraus and A.~Shah,
  arXiv:1012.5072 [hep-th].
}

\lref\FreedmanGP{
  D.~Z.~Freedman, S.~S.~Gubser, K.~Pilch and N.~P.~Warner,
  ``Renormalization group flows from holography supersymmetry and a c
  theorem,''
  Adv.\ Theor.\ Math.\ Phys.\  {\bf 3}, 363 (1999)
  [arXiv:hep-th/9904017].
}

\lref\BilalCF{
  A.~Bilal,
  ``W algebras from Chern-Simons theory,''
  Phys.\ Lett.\  B {\bf 267}, 487 (1991).
}

\lref\PopeCR{
  C.~N.~Pope and K.~S.~Stelle,
  ``SU(infinity), SU+(infinity) AND AREA PRESERVING ALGEBRAS,''
  Phys.\ Lett.\  B {\bf 226}, 257 (1989).
}

\lref\BergshoeffNS{
  E.~Bergshoeff, M.~P.~Blencowe and K.~S.~Stelle,
  ``AREA PRESERVING DIFFEOMORPHISMS AND HIGHER SPIN ALGEBRA,''
  Commun.\ Math.\ Phys.\  {\bf 128}, 213 (1990).
}

\lref\CampoleoniHG{
  A.~Campoleoni, S.~Fredenhagen and S.~Pfenninger,
  ``Asymptotic W-symmetries in three-dimensional higher-spin gauge theories,''
  arXiv:1107.0290 [hep-th].
}

\lref\HullKF{
  C.~M.~Hull,
  ``Lectures on W gravity, W geometry and W strings,''
  arXiv:hep-th/9302110.
}

\lref\EguchiSB{
  T.~Eguchi and H.~Ooguri,
  ``Conformal and Current Algebras on General Riemann Surface,''
  Nucl.\ Phys.\  B {\bf 282}, 308 (1987).
}

\lref\BergshoeffYD{
  E.~Bergshoeff, C.~N.~Pope, L.~J.~Romans, E.~Sezgin and X.~Shen,
  ``THE SUPER W(infinity) ALGEBRA,''
  Phys.\ Lett.\  B {\bf 245}, 447 (1990).
}

\lref\PopeKC{
  C.~N.~Pope, L.~J.~Romans and X.~Shen,
  ``A NEW HIGHER SPIN ALGEBRA AND THE LONE STAR PRODUCT,''
  Phys.\ Lett.\  B {\bf 242}, 401 (1990).}

\lref\WittenHC{
  E.~Witten,
  ``(2+1)-Dimensional Gravity as an Exactly Soluble System,''
  Nucl.\ Phys.\  B {\bf 311}, 46 (1988).
}

\lref\AchucarroVZ{
  A.~Achucarro and P.~K.~Townsend,
  ``A Chern-Simons Action for Three-Dimensional anti-De Sitter Supergravity
  Theories,''
  Phys.\ Lett.\  B {\bf 180}, 89 (1986).
}

\lref\BrownNW{
  J.~D.~Brown and M.~Henneaux,
  ``Central Charges in the Canonical Realization of Asymptotic Symmetries: An
  Example from Three-Dimensional Gravity,''
  Commun.\ Math.\ Phys.\  {\bf 104}, 207 (1986).
}


\lref\HaggiManiRU{
  P.~Haggi-Mani and B.~Sundborg,
  ``Free large N supersymmetric Yang-Mills theory as a string theory,''
  JHEP {\bf 0004}, 031 (2000)
  [arXiv:hep-th/0002189].
}

\lref\KonsteinBI{
  S.~E.~Konstein, M.~A.~Vasiliev and V.~N.~Zaikin,
  ``Conformal higher spin currents in any dimension and AdS/CFT
  correspondence,''
  JHEP {\bf 0012}, 018 (2000)
  [arXiv:hep-th/0010239].
}

\lref\SundborgWP{
  B.~Sundborg,
  ``Stringy gravity, interacting tensionless strings and massless higher
  spins,''
  Nucl.\ Phys.\ Proc.\ Suppl.\  {\bf 102}, 113 (2001)
  [arXiv:hep-th/0103247].
}

\lref\MikhailovBP{
  A.~Mikhailov,
  ``Notes on higher spin symmetries,''
  arXiv:hep-th/0201019.
}

\lref\SezginRT{
  E.~Sezgin and P.~Sundell,
  ``Massless higher spins and holography,''
  Nucl.\ Phys.\  B {\bf 644}, 303 (2002)
  [Erratum-ibid.\  B {\bf 660}, 403 (2003)]
  [arXiv:hep-th/0205131].
}

\lref\KlebanovJA{
  I.~R.~Klebanov and A.~M.~Polyakov,
  ``AdS dual of the critical O(N) vector model,''
  Phys.\ Lett.\  B {\bf 550}, 213 (2002)
  [arXiv:hep-th/0210114].
}

\lref\GiombiWH{
  S.~Giombi and X.~Yin,
  ``Higher Spin Gauge Theory and Holography: The Three-Point Functions,''
  JHEP {\bf 1009}, 115 (2010)
  [arXiv:0912.3462 [hep-th]].
}

\lref\GiombiVG{
  S.~Giombi and X.~Yin,
  ``Higher Spins in AdS and Twistorial Holography,''$\quad \quad$
  arXiv:1004.3736 [hep-th].
}

\lref\HenneauxXG{
  M.~Henneaux and S.~J.~Rey,
  ``Nonlinear W(infinity) Algebra as Asymptotic Symmetry of Three-Dimensional
  Higher Spin Anti-de Sitter Gravity,''
  JHEP {\bf 1012}, 007 (2010)
  [arXiv:1008.4579 [hep-th]].
}

\lref\CampoleoniZQ{
  A.~Campoleoni, S.~Fredenhagen, S.~Pfenninger and S.~Theisen,
  ``Asymptotic symmetries of three-dimensional gravity coupled to higher-spin
  fields,''
  JHEP {\bf 1011}, 007 (2010)
  [arXiv:1008.4744 [hep-th]].
}

\lref\GaberdielAR{
  M.~R.~Gaberdiel, R.~Gopakumar and A.~Saha,
  ``Quantum W-symmetry in AdS$_3$,''
  JHEP {\bf 1102}, 004 (2011)
  [arXiv:1009.6087 [hep-th]].
}

\lref\GaberdielPZ{
  M.~R.~Gaberdiel and R.~Gopakumar,
  ``An AdS$_3$ Dual for Minimal Model CFTs,''
  arXiv:1011.2986 [hep-th].
}

\lref\DouglasRC{
  M.~R.~Douglas, L.~Mazzucato and S.~S.~Razamat,
  ``Holographic dual of free field theory,''
  arXiv:1011.4926 [hep-th].
}

\lref\CastroCE{
  A.~Castro, A.~Lepage-Jutier and A.~Maloney,
  ``Higher Spin Theories in AdS$_3$ and a Gravitational Exclusion Principle,''
  JHEP {\bf 1101}, 142 (2011)
  [arXiv:1012.0598 [hep-th]].
}

\lref\GaberdielWB{
  M.~R.~Gaberdiel and T.~Hartman,
  ``Symmetries of Holographic Minimal Models,''
  arXiv:1101.2910 [hep-th].
}

\lref\FradkinKS{
  E.~S.~Fradkin and M.~A.~Vasiliev,
  ``On the Gravitational Interaction of Massless Higher Spin Fields,''
  Phys.\ Lett.\  B {\bf 189}, 89 (1987).
}

\lref\FradkinQY{
  E.~S.~Fradkin and M.~A.~Vasiliev,
  ``Cubic Interaction in Extended Theories of Massless Higher Spin Fields,''
  Nucl.\ Phys.\  B {\bf 291}, 141 (1987).
}

\lref\VasilievTK{
  M.~A.~Vasiliev,
  ``Equations of motion for interacting massless fields of all spins in
  (3+1)-dimensions,''
{\it  In *Moscow 1990, Proceedings, Symmetries and algebraic structures in physics, pt. 1* 15-33}
}

\lref\VasilievAV{
  M.~A.~Vasiliev,
  ``More On Equations Of Motion For Interacting Massless Fields Of All Spins In
  (3+1)-Dimensions,''
  Phys.\ Lett.\  B {\bf 285}, 225 (1992).
}

\lref\BlencoweGJ{
  M.~P.~Blencowe,
  ``A Consistent Interacting Massless Higher Spin Field Theory In D = (2+1),''
  Class.\ Quant.\ Grav.\  {\bf 6}, 443 (1989).
}

\lref\BergshoeffNS{
  E.~Bergshoeff, M.~P.~Blencowe and K.~S.~Stelle,
  ``Area Preserving Diffeomorphisms And Higher Spin Algebra,''
  Commun.\ Math.\ Phys.\  {\bf 128}, 213 (1990).
}

\lref\ZamolodchikovWN{
  A.~B.~Zamolodchikov,
  ``Infinite Additional Symmetries In Two-Dimensional Conformal Quantum Field
  Theory,''
  Theor.\ Math.\ Phys.\  {\bf 65}, 1205 (1985)
  [Teor.\ Mat.\ Fiz.\  {\bf 65}, 347 (1985)].
}

\lref\KrausWN{
  P.~Kraus,
  ``Lectures on black holes and the AdS(3)/CFT(2) correspondence,''
  Lect.\ Notes Phys.\  {\bf 755}, 193 (2008)
  [arXiv:hep-th/0609074].
}

\lref\FronsdalRB{
  C.~Fronsdal,
  ``Massless Fields With Integer Spin,''
  Phys.\ Rev.\  D {\bf 18}, 3624 (1978).
}

\lref\DidenkoTD{
  V.~E.~Didenko and M.~A.~Vasiliev,
  ``Static BPS black hole in 4d higher-spin gauge theory,''
  Phys.\ Lett.\  B {\bf 682}, 305 (2009)
  [arXiv:0906.3898 []].
}

\lref\KochCY{
  R.~d.~M.~Koch, A.~Jevicki, K.~Jin and J.~P.~Rodrigues,
  ``$AdS_4/CFT_3$ Construction from Collective Fields,''
  Phys.\ Rev.\  D {\bf 83}, 025006 (2011)
  [arXiv:1008.0633].
}


\lref\GutperleKF{
  M.~Gutperle and P.~Kraus,
  ``Higher Spin Black Holes,''
  JHEP {\bf 1105}, 022 (2011)
  [arXiv:1103.4304 [hep-th]].
}

\lref\AhnPV{
  C.~Ahn,
  ``The Large N 't Hooft Limit of Coset Minimal Models,''
  arXiv:1106.0351 [hep-th].
}

\lref\GaberdielZW{
  M.~R.~Gaberdiel, R.~Gopakumar, T.~Hartman and S.~Raju,
  ``Partition Functions of Holographic Minimal Models,''
  arXiv:1106.1897 [hep-th].
}

\lref\BershadskyBG{
  M.~Bershadsky,
  ``Conformal field theories via Hamiltonian reduction,''
  Commun.\ Math.\ Phys.\  {\bf 139}, 71 (1991).
}

\lref\PolyakovDM{
  A.~M.~Polyakov,
  ``Gauge Transformations and Diffeomorphisms,''
  Int.\ J.\ Mod.\ Phys.\  A {\bf 5}, 833 (1990).
}

\lref\BilalCF{
  A.~Bilal,
  ``W algebras from Chern-Simons theory,''
  Phys.\ Lett.\  B {\bf 267}, 487 (1991).
}

\lref\DynkinUM{
  E.~B.~Dynkin,
  ``Semisimple subalgebras of semisimple Lie algebras,''
  Trans.\ Am.\ Math.\ Soc.\  {\bf 6}, 111 (1957).
}

\lref\BaisBS{
  F.~A.~Bais, T.~Tjin and P.~van Driel,
  ``Covariantly coupled chiral algebras,''
  Nucl.\ Phys.\  B {\bf 357}, 632 (1991).
}

\lref\ForgacsAC{
  P.~Forgacs, A.~Wipf, J.~Balog, L.~Feher and L.~O'Raifeartaigh,
  ``Liouville and Toda Theories as Conformally Reduced WZNW Theories,''
  Phys.\ Lett.\  B {\bf 227}, 214 (1989).
}

\lref\deBoerIZ{
  J.~de Boer and T.~Tjin,
  ``The Relation between quantum W algebras and Lie algebras,''
  Commun.\ Math.\ Phys.\  {\bf 160}, 317 (1994)
  [arXiv:hep-th/9302006].
}

\lref\ChangMZ{
  C.~M.~Chang and X.~Yin,
  ``Higher Spin Gravity with Matter in $AdS_3$ and Its CFT Dual,''
  arXiv:1106.2580 [hep-th].
}

\lref\MaldacenaRE{
  J.~M.~Maldacena,
  ``The Large N limit of superconformal field theories and supergravity,''
  Adv.\ Theor.\ Math.\ Phys.\  {\bf 2}, 231 (1998)
  [Int.\ J.\ Theor.\ Phys.\  {\bf 38}, 1113 (1999)]
  [arXiv:hep-th/9711200].
}

\lref\WittenQJ{
  E.~Witten,
  ``Anti-de Sitter space and holography,''
  Adv.\ Theor.\ Math.\ Phys.\  {\bf 2}, 253 (1998)
  [arXiv:hep-th/9802150].
}

\lref\GubserBC{
  S.~S.~Gubser, I.~R.~Klebanov and A.~M.~Polyakov,
  ``Gauge theory correlators from noncritical string theory,''
  Phys.\ Lett.\  B {\bf 428}, 105 (1998)
  [arXiv:hep-th/9802109].
}

\lref\JevickiSS{
  A.~Jevicki, K.~Jin and Q.~Ye,
  ``Collective Dipole Model of AdS/CFT and Higher Spin Gravity,''
  arXiv:1106.3983 [hep-th].
}

\lref\WaldNT{
  R.~M.~Wald,
  ``Black hole entropy is the Noether charge,''
  Phys.\ Rev.\  D {\bf 48}, 3427 (1993)
  [arXiv:gr-qc/9307038].
}

\lref\ProkushkinBQ{
  S.~F.~Prokushkin and M.~A.~Vasiliev,
  ``Higher spin gauge interactions for massive matter fields in 3-D AdS
  space-time,''
  Nucl.\ Phys.\  B {\bf 545}, 385 (1999)
  [arXiv:hep-th/9806236].
}

\lref\ProkushkinVN{
  S.~Prokushkin and M.~A.~Vasiliev,
  ``3-d higher spin gauge theories with matter,''
  arXiv:hep-th/9812242.
}

\lref\tmg{
  B.~Chen, J.~Long and J.~b.~Wu,
  ``Spin-3 Topological Massive Gravity,''
  arXiv:1106.5141 [hep-th].}

\lref\tmgg{
  A.~Bagchi, S.~Lal, A.~Saha and B.~Sahoo,
  ``Topologically Massive Higher Spin Gravity,''
  arXiv:1107.0915 [hep-th].}

\lref\gab{
  M.~R.~Gaberdiel and T.~Hartman,
  ``Symmetries of Holographic Minimal Models,''
  JHEP {\bf 1105}, 031 (2011)
  [arXiv:1101.2910 [hep-th]].}

\lref\cfp{
  A.~Campoleoni, S.~Fredenhagen and S.~Pfenninger,
 ``Asymptotic W-symmetries in three-dimensional higher-spin gauge theories,''
  arXiv:1107.0290 [hep-th].}

\lref\prs{
  C.~N.~Pope, L.~J.~Romans and X.~Shen,
  ``W(infinity) AND THE RACAH-WIGNER ALGEBRA,''
  Nucl.\ Phys.\  B {\bf 339}, 191 (1990).}

\lref\cacc{
  S.~L.~Cacciatori, M.~M.~Caldarelli, A.~Giacomini, D.~Klemm and D.~S.~Mansi,
  ``Chern-Simons formulation of three-dimensional gravity with torsion and
  nonmetricity,''
  J.\ Geom.\ Phys.\  {\bf 56}, 2523 (2006)
  [arXiv:hep-th/0507200].}

\lref\VasilievBA{
  M.~A.~Vasiliev,
  ``Higher spin gauge theories: Star product and AdS space,''
  arXiv:hep-th/9910096.
}

\lref\BekaertVH{
  X.~Bekaert, S.~Cnockaert, C.~Iazeolla and M.~A.~Vasiliev,
  ``Nonlinear higher spin theories in various dimensions,''
  arXiv:hep-th/0503128.
}

\lref\SundborgWP{
  B.~Sundborg,
  ``Stringy gravity, interacting tensionless strings and massless higher
  spins,''
  Nucl.\ Phys.\ Proc.\ Suppl.\  {\bf 102}, 113 (2001)
  [arXiv:hep-th/0103247].
}

\lref\DasVW{
  S.~R.~Das and A.~Jevicki,
  ``Large N collective fields and holography,''
  Phys.\ Rev.\  D {\bf 68}, 044011 (2003)
  [arXiv:hep-th/0304093].
}

\lref\ChangMZ{
  C.~M.~Chang and X.~Yin,
  ``Higher Spin Gravity with Matter in AdS$_3$ and Its CFT Dual,''
  arXiv:1106.2580 [hep-th].
}

\lref\AmmonNK{
  M.~Ammon, M.~Gutperle, P.~Kraus and E.~Perlmutter,
  ``Spacetime Geometry in Higher Spin Gravity,''
  arXiv:1106.4788 [hep-th].
}

\lref\StromingerEQ{
  A.~Strominger,
  ``Black hole entropy from near horizon microstates,''
  JHEP {\bf 9802}, 009 (1998)
  [arXiv:hep-th/9712251].
}

\lref\BanadosWN{
  M.~Banados, C.~Teitelboim and J.~Zanelli,
  ``The Black hole in three-dimensional space-time,''
  Phys.\ Rev.\ Lett.\  {\bf 69}, 1849 (1992)
  [arXiv:hep-th/9204099].
}

\lref\CampoleoniHG{
  A.~Campoleoni, S.~Fredenhagen and S.~Pfenninger,
  arXiv:1107.0290 [hep-th].
}

\lref\WaldNT{
  R.~M.~Wald,
  ``Black hole entropy is the Noether charge,''
  Phys.\ Rev.\  D {\bf 48}, 3427 (1993)
  [arXiv:gr-qc/9307038].
}

\lref\GibbonsUE{
  G.~W.~Gibbons and S.~W.~Hawking,
  ``Action Integrals and Partition Functions in Quantum Gravity,''
  Phys.\ Rev.\  D {\bf 15}, 2752 (1977).
}

\lref\PopeEW{
  C.~N.~Pope, L.~J.~Romans and X.~Shen,
  ``The Complete Structure of W(Infinity),''
  Phys.\ Lett.\  B {\bf 236}, 173 (1990).
}

\lref\BakasRY{
  I.~Bakas and E.~Kiritsis,
  ``BOSONIC REALIZATION OF A UNIVERSAL W ALGEBRA AND Z(infinity)
  PARAFERMIONS,''
  Nucl.\ Phys.\  B {\bf 343}, 185 (1990)
  [Erratum-ibid.\  B {\bf 350}, 512 (1991)].
}

\lref\HullSA{
  C.~M.~Hull,
  ``W gravity anomalies 1: Induced quantum W gravity,''
  Nucl.\ Phys.\  B {\bf 367}, 731 (1991).
}

\lref\HullKF{
  C.~M.~Hull,
  ``Lectures on W gravity, W geometry and W strings,''
  arXiv:hep-th/9302110.
}

\lref\EguchiSB{
  T.~Eguchi and H.~Ooguri,
  ``Conformal and Current Algebras on General Riemann Surface,''
  Nucl.\ Phys.\  B {\bf 282}, 308 (1987).
}

\lref\Iazeolla{
  C.~Iazeolla, P.~Sundell,
  ``Families of exact solutions to Vasiliev's 4D equations with spherical, cylindrical and biaxial symmetry,''
  
  [arXiv:1107.1217 [hep-th]].}


\Title{\vbox{\baselineskip14pt
}} {\vbox{\centerline {Partition functions of higher spin black holes }
\medskip\vbox{\centerline {and their CFT duals}}}}
\centerline{Per
Kraus and Eric Perlmutter\foot{pkraus@ucla.edu, perl@physics.ucla.edu}}
\bigskip
\centerline{\it{Department of Physics and Astronomy}}
\centerline{${}$\it{University of California, Los Angeles, CA 90095, USA}}

\baselineskip14pt

\vskip .3in

\centerline{\bf Abstract}
\vskip.2cm

We find  black hole solutions of   $D=3$ higher-spin gravity in the \hsl~$\oplus$~\hsl\ Chern-Simons formulation.
These solutions have a  spin-3 chemical potential, and carry
nonzero values for an infinite number of charges of the asymptotic
$\mw_{\infty}[\l]$ symmetry.  Applying a previously developed set of rules for ensuring  smooth solutions, we compute the black hole partition function perturbatively in the chemical potential.
At $\lambda =0, 1$ we compare our result against boundary CFT computations involving free bosons and fermions, and find perfect agreement.
For generic $\lambda$ we expect that our gravity result will
match the partition function of the coset CFTs conjectured by Gaberdiel and Gopakumar to be dual to these bulk theories.

\Date{August  2011}
\baselineskip13pt


\newsec{Introduction}

Higher spin theories of gravity, as developed by Vasiliev and collaborators, are fascinating theories that lie, in some sense, halfway
between ordinary gravity and  string theory; see \refs{\VasilievBA,\BekaertVH} for reviews.    In particular, while they share with string theories such features as infinite towers of higher spin fields and nonlocal dynamics, their full (classical) equations of motion can be written down in a background independent manner.  Relatively recently, it has been realized that higher spin theories may lead to soluble examples of the AdS/CFT correspondence \refs{\HaggiManiRU,\SundborgWP,\SezginRT,
\KlebanovJA,\DasVW,\GiombiWH,\GiombiVG,\KochCY,\DouglasRC,\JevickiSS,
\GaberdielPZ,\GaberdielWB,\GaberdielZW,\ChangMZ}.

Given the major role that black holes play in attempts to understand quantum gravity and holography, we hardly need to justify the motivation for constructing and studying black holes in higher spin theories.
Here, extending previous work \refs{\GutperleKF,\AmmonNK}, we focus on the D=2+1 dimensional higher spin theory \refs{\BergshoeffNS,\BlencoweGJ,\ProkushkinBQ}, and we will
consider black holes that generalize the BTZ solution \BanadosWN; see \refs{\DidenkoTD,\CastroCE, \Iazeolla} for other work on black holes in higher spin gravity.

 In ordinary
gravity, the asymptotic symmetry algebra consists of left and right moving Virasoro algebras \BrownNW, and the general rotating BTZ black hole carries independent left and right moving Virasoro zero mode charges.   The BTZ entropy takes the form of Cardy's formula, and so matches elegantly with the entropy of any CFT dual with the same central charge \StromingerEQ.

In higher spin gravity, the asymptotic symmetry algebra contains one additional (left and right moving) conserved  charge for each higher spin field \refs{\GaberdielWB, \HenneauxXG,\CampoleoniZQ,\CampoleoniHG}.   We expect there to exist black holes that carry these conserved charges.   If these black holes can be found and their entropy computed, we can use these results to test any proposed AdS/CFT duality involving these theories,
assuming that the entropy can be computed on the CFT side as well.

Apart from AdS/CFT applications, understanding how to compute the entropy of higher spin black holes represents an interesting technical and conceptual challenge: due to the nonstandard form of these theories, one cannot apply (at least with current understanding) such familiar approaches as the area law, the Wald entropy \WaldNT, or the Gibbons-Hawking Euclidean action  \GibbonsUE.  This problem was studied in \refs{\GutperleKF,\AmmonNK} in the
simplest version of higher spin gravity, based on SL(3,R)$\times$ SL(3,R)
Chern-Simons theory.  Black hole solutions were found and their entropy computed by appealing to first principles, in particular to the fact that what constitutes a physically satisfactory entropy is that it appear
correctly in a thermodynamical first law.   The other novel ingredient, reviewed in more detail below, was to give a gauge invariant characterization of a smooth event horizon, based on the holonomies of the Chern-Simons connection.   The usual approach of determining the Hawking temperature
by compactifying imaginary time and demanding the absence of a conical singularity is not straightforward to apply in this context, as it is not fully gauge invariant.  Indeed, as was shown  explicitly in \AmmonNK, even the existence
of an event horizon in the metric is a gauge dependent statement: a gauge transformation was exhibited that transformed the metric between a black hole and a traversable  wormhole.  Despite all these subtleties, by following the logic in \refs{\GutperleKF,\AmmonNK} it is possible to unambiguously compute all physical
properties of these higher spin black holes.

In this paper we apply our previous logic to the Vasiliev theories containing an infinite tower of higher spin fields \ProkushkinBQ.   These theories are
based on a one-parameter family of infinite dimensional gauge algebras,
denoted hs$[\lambda]$ \prs.  The BTZ black hole is still a solution of this theory, but rather too simple as it carries vanishing values for all higher spin charges.   To access the higher spin sector we turn on a nonzero spin-3 chemical potential, $\alpha$.   Due to the nonlinear structure of the theory, this triggers nonzero values for the entire infinite tower of higher spin charges.  The values of these charges, and the full smooth
solution, can be determined systematically using perturbation theory in
$\alpha$.   As in the SL(3,R) case studied in \refs{\GutperleKF,\AmmonNK}, crucial input is provided by demanding a gauge invariant smooth horizon, as expressed in terms of the holonomies.
The main output of this procedure is a result for the black hole partition function, $ Z(\tau, \alpha; \taub, \overline{\alpha})$.   Here $\tau$ is the modular parameter of the torus that describes the boundary of the Euclidean black hole geometry, and $\alpha$ is the leftmoving spin-3 chemical potential, as noted above. Similarly, $\overline{\alpha}$ is the rightmoving analog of $\alpha$.
As usual, given the partition function, other thermodynamical quantities such as the energy and entropy can be obtained by suitable differentiation.   Our result for the black hole partition function, up to order $\alpha^8$, is
\eqn\ina{\eqalign{ \ln Z & =
{i \pi k \over 2\tau} \Bigg[1 -{ 4\over 3}~{\alpha^2 \over \tau^4}
+{400 \over 27}~ { \lambda^2-7 \over \lambda^2-4}~{\alpha^4\over \tau^8} -{1600 \over 27}~{5\lambda^4 -85 \lambda^2 +377 \over (\lambda^2-4)^2}~ {\alpha^6 \over \tau^{12}}\cr
&\quad\quad\quad~   +{32000\over 81}~{20\l^6-600\l^4+6387\l^2-23357\over(\l^2-4)^3}~{\a^8\over\t^{16}}\Bigg]+ \ldots \cr
&  \quad +{\rm rightmoving}  }}
where the rightmoving part is obtained by replacing $\tau$ and
$\alpha$ by $\taub$ and $\overline{\alpha}$ in the obvious way.
The leading term in \ina\ is the usual BTZ result.  We note that the factor of $\lambda^2-4$ appearing in the denominators is just due to our normalization convention for the spin-3 charge, and has no special significance. The entropy formula obtained from this partition function can be thought of as a generalized version of Cardy's formula to include higher spin charge.

As we now discuss, this result can be used to test the  AdS/CFT duality conjectured recently by Gaberdiel and Gopakumar \GaberdielPZ.  In particular, they propose to consider the $\Wc_N$ minimal model  coset CFT
\eqn\inb{ {SU(N)_k \oplus SU(N)_1 \over SU(N)_{k+1}} }
The 't Hooft limit is defined as
\eqn\inc{ N, k \rt \infty~,\quad    \lambda \equiv {N \over k+N}~~~{\rm fixed} }
Gaberdiel and Gopakumar conjecture that this theory in the 't Hooft limit is dual to the bulk higher spin theory based on the algebra hs$[\lambda]$, along with some additional scalar fields that will play no role in the present discussion.    Evidence for this proposal, based on symmetries, RG flows,  and the perturbative spectrum, is discussed in \refs{\GaberdielPZ,\GaberdielWB,\GaberdielZW,\ChangMZ}.

To make contact with our black hole result, we should consider the partition function of this theory with the insertion of a spin-3 chemical potential,\foot{In general, one might wonder whether such traces are convergent.  In the following we will be considering perturbation theory in $\alpha$ and $\alphab$, and such issues will not arise.}
\eqn\inc{  Z_{CFT}(\tau, \alpha; \taub, \overline{\alpha}) = \Tr \Big[ e^{4\pi^2 i( \tau \hat{\Lc}+ \alpha \hat{\Wc}  -\taub \hat{\overline{\Lc}} -\overline{\alpha} \hat{\overline{\Wc}})}\Big] }
where the operators denote the suitably normalized Virasoro and spin-3 zero modes.  As is standard, to compare with the black hole side we should consider the leading high temperature asymptotics, defined here by taking $\tau, \alpha \rt 0$ with $\alpha/\tau^2$ fixed. This is a version of the Cardy limit, generalized to include the higher spin chemical potential. In this limit, the duality conjecture asserts that \ina\ and \inc\ should agree.

While it should be possible to test this prediction for general $\lambda$, in this paper we will only carry out the CFT computation for the special values $\lambda =0, 1$.  The reason why these values are more tractable is as follows.   In general, the symmetry algebra controlling the coset theory in the 't Hooft limit is believed to be the infinite dimensional algebra $\Wc_{\infty}[\lambda]$. This needs to be so in order for the duality conjecture to be true --- for instance, the $\Wc_{\infty}[\lambda]$ algebra is the asymptotic symmetry algebra of \hsl\ gravity on AdS$_3$ \refs{\GaberdielWB,\HenneauxXG,\CampoleoniZQ,\CampoleoniHG} --- but  independent evidence is also available \refs{\GaberdielWB,\GaberdielZW}. At $\lambda = 0, 1$
these algebras simplify.   At $\lambda =1$, after a change of basis, the algebra turns into the {\it linear} algebra $\Wc_{\infty}^{\rm PRS}$ \PopeEW.
Importantly for us, this algebra can be represented in terms of a collection of free bosons, with the higher spin currents being quadratic in the bosons \refs{\BakasRY,\HullSA,\HullKF}. Since the bosons are free, we can of course compute \inc\ exactly for this theory.
If we make the plausible assumption (justified in more detail in the text) that this free boson theory should share the same high temperature partition function as the coset theory at $\lambda =1$, then we arrive at the
striking prediction that our black hole result should match a certain free boson partition function.  Up to the order that we have checked, this turns out to be correct: we find precise agreement with \ina\ at $\lambda =1$!

An analogous story holds at $\lambda =0$, but now in terms of free fermions.   At $\lambda =0$ the $\Wc_\infty[\lambda]$ algebra is related to the algebra $\Wc_{1+\infty}$ \PopeKC\ by a constraint that removes the spin-1 current.   Since the $\Wc_{1+\infty}$ algebra can be represented by free fermions \BergshoeffYD, we can compute its partition function with the spin-1 constraint imposed.  We then find precise agreement with \ina\ at $\lambda =0$.

We view these results as providing strong evidence  for the validity of our  rules for treating black holes in higher spin gravity, and for applying them to the  conjecture of Gaberdiel and Gopakumar.  Further tests along these lines are clearly possible, most obviously by extending the CFT computations to generic $\lambda$, and pushing the comparison to higher (ideally all) orders in $\alpha$.  Another useful generalization would be to turn on additional chemical potentials.
More ambitiously, it seems reasonable to hope that these comparisons will lead to a deeper understanding of how the duality is working at a fundamental level.

The remainder of this paper is organized as follows.  In section 2, after reviewing the Chern-Simons formulation of higher spin gravity, we present the rules for constructing higher spin black hole solutions. These rules are applied to the hs$[\lambda]$ theories in section 3, and the black hole partition function is computed.  In section 4 we compute the partition functions for free bosons and fermions, and demonstrate agreement with
the bulk result for $\lambda =0,1$.  Section 5 contains a discussion of the
implications of our results for the AdS/CFT correspondence.  Appendix A
gives the hs$[\lambda]$ structure constants, and in appendix B we display
certain holonomy equations in detail.

\newsec{Constructing black holes in higher-spin gravity}

We begin with a  review of the Chern-Simons formulation of D=2+1 gravity, along with the rules developed in \refs{\GutperleKF,\AmmonNK} for constructing black hole solutions.  These rules apply to any Chern-Simons formulation of gravity in which the connections take values in a Lie algebra that contains SL(2,R).

\subsec{Chern-Simons gravity}

It was discovered over two decades ago that Einstein gravity with a negative cosmological constant can be re-written as a \slt~$\times$~\slt\ Chern-Simons theory \refs{\AchucarroVZ,\WittenHC}
. With 1-forms $(A,\bar{A})$ taking values in the Lie algebra of \slt, the action is
\eqn\abb{ S = S_{CS}[A] - S_{CS}[\Ab]}
where
\eqn\ac{  S_{CS}[A] ={k\over 4\pi} \int\! \Tr \left(A\wedge dA +{2\over 3} A\wedge A \wedge A\right)}
The Chern-Simons level $k$ is related to the
Newton constant $G$ and AdS$_3$ radius $l$ as
\eqn\ad{k ={l\over 4G}}

The Chern-Simons equations of motion correspond
to vanishing field strengths,
\eqn\ae{ F = dA + A\wedge A =0~,\quad \Fb= d\Ab+\Ab\wedge \Ab =0 }

We can consider taking $(A,\Ab)$ to lie in some other Lie algebra besides \slt, which we will denote $\cg$. Doing so is equivalent to coupling some set of higher spin fields to Einstein gravity, where the rank of $\cg$ determines the number of higher spin fields. Taking $\cg=$~\sln, for example, one has a theory of Einstein gravity coupled to a tower of symmetric tensor fields of spins $s=3,4,\ldots,N$. Taking $\cg$ to be an algebra of infinite rank introduces an infinite tower of such spins; the details of the theory depend on  which algebra one chooses. In all of these cases, one recovers Einstein gravity upon restriction to a \slt~$\times$~\slt\ subalgebra of $\cg \times \cg$.  Note that in general
a given $\cg$ admits many inequivalent embeddings of SL(2,R); this leads to the appearance of multiple AdS$_3$ vacua in the theory \AmmonNK.

\vs

In this paper, we will take $\cg$ to be the infinite-dimensional higher spin algebra \hsl, which appears in higher spin contexts old \refs{\ProkushkinBQ,\PopeEW} and new \refs{\GaberdielPZ,\GaberdielWB,\HenneauxXG,\CampoleoniZQ,\CampoleoniHG}. Our goal will be to construct a black hole solution in such a theory with  higher spin charges turned on. In preparation, we review  the spin-3 black hole \paps\ of the $\cg=$~\slth\ theory, which possesses only the spin-3 field in addition to the graviton.
\subsec{Review: the spin-3 black hole}
The spacetime interpretation and asymptotic symmetries of the \slth~$\times$~\slth\ Chern-Simons theory were treated in  detail in \CampoleoniZQ, following which \paps\ discovered smooth, asymptotically AdS$_3$, black hole solutions with nonzero spin-3 charge and consistent thermodynamics. We refer to those papers for details, here extracting only the essential lessons about how to make a spin-3 black hole.

We start from the simple fact that the ordinary BTZ black hole is a solution of the \slth~$\times$~\slth\ theory. Denoting a \slt\ subalgebra of generators as $\{L_{\pm1},L_0\}$, the BTZ solution in Chern-Simons language is
\eqn\ah{\eqalign{ A&= \Big(e^\rho L_1 -{2\pi \over k}\Lc  e^{-\rho}L_{-1}\Big)dx^+ +L_0 d\rho \cr
 \Ab&= -\Big(e^\rho L_{-1} -{2\pi \over k}\Lcb e^{-\rho} L_{1}\Big)dx^- -L_0 d\rho \cr }}
Here $(\rho, x^\pm)$ are the spacetime coordinates, with $x^\pm = t\pm \phi$.  $(\ml,\Lcb)$ are the conserved charges carried by the black hole,  {\it i.e.}  linear combinations of the mass and angular momentum.  To write down the metric $g_{\mu\nu}$ and spin-3 field $\varphi_{\mu\nu\gamma}$, we introduce a generalized vielbein $e$ and spin connection $\omega$ as
\eqn\af{ A = \omega+e ~,\quad \Ab= \omega -e }
Expanding $e$ and $\omega$ in a basis of 1-forms $dx^\mu$,
the spacetime fields are identified as
\eqn\ag{ g_{\mu\nu} = {1\over 2} \Tr (e_\mu e_{\nu} )~,\quad \varphi _{\mu\nu\gamma} ={1\over 3! } \Tr ( e_{(\mu} e_\nu e_{\gamma)})}
Of course, for the BTZ solution one has $\varphi_{\mu\nu\gamma}=0$.

The Euclidean BTZ black is obtained by taking $dx^+ = dz$ and $dx^- = -d\zb$.  To avoid a conical singularity at the horizon we need to make the identification
$(z,\zb)\cong (z+2\pi \tau, \zb+ 2\pi \taub)$, with
\eqn\ak{\ml=-{k\over8\pi\t^2}~, \quad \Lcb=-{k\over8\pi\bar{\t}^2}}
The inverse temperature of this solution, $\beta$, and angular velocity of the horizon, $\Omega$, are then given by
\eqn\akk{\t={i\beta+i\beta\Omega\over2\pi}~, \quad \bar{\t}={-i\beta+i\beta\Omega\over2\pi}}

Of particular interest for present purposes are the holonomies of the gauge connections around the Euclidean time circle defined above. We define
\eqn\ai{ \omega = 2\pi( \t A_+-\bar{\t}A_-)}
and similarly for the barred holonomy.  The gauge invariant information contained in $\omega$ is given by its two independent eigenvalues, or equivalently by the values of $\Tr (\omega^2)$ and $\Tr (\omega^3)$. For the BTZ solution we compute
\eqn\aj{\Tr(\omega^2)=-8\pi^2~,\quad \Tr(\omega^3)=0 }
The second of these is identically true, and the first is true by virtue of \ak.

Now we want to add spin-3 charges $(\mw,\Wcb)$.   Since black holes represent states of thermodynamic equilibrium, we of course also need to
turn on the corresponding conjugate potentials $(\mu,\bar{\mu})$. Just as in the BTZ case smoothness at the horizon fixed the relations \ak\ between
the spin-2 charges and potentials, here also we expect that smoothness will fix a relation between the spin-3 charges and their conjugate potentials.
The main subtlety, elaborated on in detail in \paps, is
that one cannot impose smoothness by naively examining the local geometry at the horizon, since this local geometry, and even the very existence of the horizon, is not SL(3,R)$\times $ SL(3,R) gauge invariant.

The primary lesson of \paps\ is that one should consider the equations \aj\ to be the gauge-invariant characterization of a smooth horizon for any solution of the \slth~$\times$~\slth\ theory. In a theory of flat connections, the holonomy captures the physics encoded in a given connection, and demanding that the time circle holonomy  take its gauge-invariant value as fixed by the BTZ metric enforces consistency via fixing smoothness at the origin of the Euclidean plane. Conveniently, this does not require passage to the metric-like fields, as in \ag: given some candidate connection for a spin-3 black hole, we fix the charges by fixing the time circle holonomy.

There are two pieces of evidence for the validity of this proposal.
As noted above, in a generic gauge the connection for a spin-3 black hole may correspond to a metric with no event horizon.   However, \papb\ showed
that if (and plausibly only if) the holonomy conditions \aj\ are satisfied,
somewhere on the gauge orbit of this connection lies a metric with a completely smooth event horizon.\foot{Strictly speaking, this was only shown in the case of static solutions.}  This is the sense in which we can meaningfully refer to such solutions as black holes.

Furthermore, the holonomy prescription guarantees a sensible thermodynamics. Before enforcing these conditions, the solution is a function of the chemical potential $\mu$, the spin-3 charge $\mw$, the leftmoving momentum $\ml$ and the inverse temperature $\t$, along with all barred partners. We want to think of the black hole as a saddle point contribution to a partition function of the form
\eqn\am{Z(\tau,\alpha; \bar{\t},\bar{\a})= \Tr [e^{4\pi^2 i (\tau \Lc + \alpha \Wc-\bar{\t}\Lcb-\bar{\a}\Wcb)}]}
Focus on the unbarred quantities. Since \am\ implies $\Lc \sim {\p Z\over \p \tau}$ and $\Wc\sim {\p Z \over \p \alpha}$, $Z(\t,\a;\bar{\t},\bar{\a})$ will exist only if the charge assignments obey the integrability condition
\eqn\al{{\p \Lc \over \p \alpha} = {\p \Wc \over \p \tau}}

Happily, one finds consistency between the holonomy conditions and integrability upon taking
\eqn\alm{\alpha = \bar{\t} \mu~, \quad \bar{\a} = \t\bar{\mu}}
The entropy derived from \am\ is now consistent with the first law of thermodynamics, and is defined without having to address the puzzles that the extended gauge-invariance of the spin-3 theory poses, in particular the poorly understood relation between entropy and any geometric quantity analogous to the black hole horizon area. \vs

\subsec{How to make  higher spin black holes}

With this example in mind, we turn to the case of a higher spin theory built upon an arbitrary Lie algebra $\cg$ which contains a \slt\ subalgebra. The prescription for building smooth black holes with higher spin charge is as follows:\vs

\item{1.} Write down a BTZ solution.

\item{2.} Compute the BTZ time circle holonomy eigenvalues.

\item{3.} Write down a flat connection that includes nonzero chemical potentials for some chosen set of higher spin charges.

\item{4.} Fix the charges in the solution by demanding that the holonomy of the solution around the time circle agrees with that of BTZ. \vs

The resulting solution will represent a black hole in the sense described above.  Note that if $\cg$ is of infinite rank, there will be an infinite number of holonomy constraints. Note also that the these solutions are not in general gauge equivalent to BTZ, since the holonomies around the angular circle will differ.

In the above algorithm, step 3 is stated the least explicitly. Fortunately, our spin-3 example suggests a  straightforward way to find the relevant connections.  To explain this, consider the explicit connection
used in \paps\foot{The generators $\{W_{\pm 2}, W_{\pm 1}, W_0\}$ transform
in the 5-dimensional representation under the adjoint action of $\{L_{\pm 1}, L_0\}$.  Also, as in \paps\ we are here using a representation of the SL(3,R) generators in terms of $3\times 3$ traceless matrices. }
\eqn\akk{\eqalign{ A&= \Big(e^\rho L_1 -{2\pi \over k}\Lc  e^{-\rho}L_{-1} -{\pi \over 2k} \Wc e^{-2\rho} W_{-2} \Big)dx^+  \cr
& \quad +\mu \Big(e^{2\rho} W_2 -{4\pi \Lc \over k}W_0 +{4\pi^2 \ml^2 \over k^2} e^{-2\rho} W_{-2} +{4\pi \Wc \over k}e^{-\rho} L_{-1}\Big)dx^-  +L_0 d\rho   }}
with an analogous formula for $\Ab$.
The corresponding metric has no event horizon, but when the holonomy conditions are obeyed it is gauge-equivalent to one that does \papb.

Written in the form \akk\ the solution appears rather complicated, but in fact the structure is quite simple.   First, following \CampoleoniZQ\ we note that  we can write
\eqn\akl{ A = b^{-1} a b + b^{-1} db }
where $b= e^{\rho L_0}$, and $a$ is obtained from $A$ by setting $\rho=d\rho =0$.   In terms of $a$, the flatness equations are simply
$[a_+, a_-]=0$.   To exhibit flatness we need only observe that
\eqn\akm{ a_- = 2\mu \left[ (a_+)^2 - {1\over 3} \Tr (a_+)^2\right]}

The form of $A_+$ corresponds to choosing the ``highest weight gauge". Namely, if we assume that $A_+$ grows as $e^{\rho}$, then by a gauge transformation it can always be put into the form in \akk\ \CampoleoniZQ.   Finally, as shown in \papa\ by a Ward identity analysis, the $\mu e^{2\rho}W_2$
term in $A_-$ gives rise to a chemical potential $\mu$ conjugate to spin-3 charge.

This discussion suggests a simple way to write down solutions that incorporate chemical potentials for higher-spin charges for any $\cg$: to turn on potentials $\mu_s$ for fields of spin $s$, simply take
\eqn\ao{\eqalign{A_+ &= A_+^{BTZ} + ({\rm higher~spin~charges})\cr
A_- &\sim \sum_s\mu_s\Big[(A_+)^{s-1}- {\rm trace}\Big]\cr
A_{\rho} &= L_0\cr}}
where  multiplication is defined by the  chosen matrix representation of the Lie algebra $\cg$. $L_0$ is the diagonal element of the \slt\ embedding into $\cg$ used to construct the BTZ solution. As usual, the terms in $A_-$ incorporate the  sources, and those in $A_+$ encode the charges. Exactly which charges one must turn on in order to have a consistent solution depends on the theory in question, and is determined by solution of the holonomy equations.

As already emphasized, the metric derived from \ao\ may not possess a horizon, but based on our study of the SL(3,R) theory we expect that there exists another connection lying on the same gauge orbit that does yield a black hole metric. Finding the explicit gauge  transformation will typically be quite involved and $\cg$-dependent, but for purposes of interpretation, we merely require its existence.
\vs

\newsec{The \mwil\ black hole}
As a last step before writing down the black hole solutions, we review the features of \hsl\ that we will need. In particular, we introduce an associative multiplication known as the ``lone-star product" \prs, the antisymmetric part of which yields the \hsl\ Lie algebra.

\subsec{\hsl\ from an associative multiplication}

The \hsl\ Lie algebra is spanned by generators labeled by a spin and a mode index. We use the notation of \GaberdielWB, in which a generator is represented as
\eqn\ba{V^s_m~, ~~ s \geq 2~, ~~ |m|<s}
The commutation relations are
\eqn\bb{[V^s_m,V^t_n] = \sum_{u=2,4,6,...}^{s+t-1}g^{st}_u(m,n;\l)V^{s+t-u}_{m+n}}
with structure constants defined in appendix A.

The generators with $s=2$ form an \slt\ subalgebra, and the remaining generators transform simply under the adjoint \slt\ action as
\eqn\be{[V^2_m,V^t_n] = (m(t-1)-n)V^t_{m+n}}
These SL(2,R) generators will be relevant in construction of the BTZ solution.

When $\l=1/2$, this algebra is isomorphic to hs(1,1), the commutator of which can be written as the antisymmetric part of the Moyal product. Similarly, the general $\l$ commutation relations \bb\ can be realized as
\eqn\bff{[V^s_m,V^t_n]= V^s_m \star V^t_n - V^t_n \star V^s_m}
if we define the associative product
\eqn\bg{V^s_m \star V^t_n \equiv \half \sum_{u=1,2,3,...}^{s+t-1}g^{st}_u(m,n;\l)V^{s+t-u}_{m+n}}
This is known as the ``lone star product'' \prs, and \bff\ follows upon using the fact that
\eqn\bh{\eqalign{g^{st}_u(m,n) &= (-1)^{u+1}g^{ts}_u(n,m)}}
The odd values of $u$ drop out of the commutator, leaving \bb. In the remainder of the paper we may resort to the shorthand
\eqn\bjj{\underbrace{\Gamma\star\Gamma\star\ldots\star\Gamma}_{N} \equiv (\Gamma)^N}
for some \hsl-valued element $\Gamma$.

Formally, $V^1_0$ is the identity element. Thus, to extract the trace from a product of generators, one picks out the $u=s+t-1$ part of \bg, up to some normalization:
\eqn\bi{\Tr (V^s_mV^t_n) \propto g^{st}_{s+t-1}(m,n;\l)\delta^{st}\delta_{m,-n}}
In order to facilitate easy comparison to the \slth\ conventions of \paps, we choose to define
\eqn\hswp{\Tr (V^s_mV^s_{-m}) = {24\over (\l^2-1)}g^{ss}_{2s-1}(m,-m;\l)}
which implies the \slt\ traces
\eqn\hswq{\Tr(V^2_1V^2_{-1})=-4~, ~~ \Tr(V^2_0V^2_{0})=2}
in agreement with the basis used in \paps.

A convenient property of the \hsl\ Lie algebra is that when $\l=N$ for integer $N\geq 2$, one can consistently set all generators with $s>N$ to zero (i.e.  factor out the ideal of the Lie algebra), and the algebra reduces to SL(N,R). This implies a similar truncation of the boundary symmetry: that is, $\mw_{\infty}[N]  =\mw_N$ upon constraining all fields of spin $s>N$ to vanish. Factoring out the ideal is automatic on the level of the trace:
\eqn\bj{\Tr(V^s_mV^s_{-m})  \propto \prod_{\sigma=2}^{s-1}(\lambda^2-\sigma^2)}
Therefore, as regards the construction of black holes, the holonomy conditions reduce to those of \sln\ when $\l=N$; this will be a useful check for us.

Another aspect of the lone star product that we wish to highlight is the following simple result for products of the highest weight SL(2,R) generator:
\eqn\bjjj{(V^2_1)^{s-1} = V^{s}_{s-1}}
A  look back at \ao\ shows that this  relation makes it easy to read off the leading behavior of $A_-$ from that of $A_+$.

\vs

In what follows, we work with a flat connection $a$ (and, implicitly, $\bar{a}$) that has no $\rho$-dependence nor $\rho$ component, as in \akm, by writing
\eqn\bl{\eqalign{A &= b^{-1}ab + b^{-1}db\cr
\Ab&= b\bar{a}b^{-1}+bdb^{-1}\cr}}
with
\eqn\bk{b = e^{\rho V^2_0}}
Conjugation  by $b$ of a generator with mode index $m$ produces a
factor $e^{m\rho}$.

\subsec{The BTZ black hole}

We now follow the prescription described in  section 2.3 for constructing the higher spin black hole.
In the \hsl\ theory, the  BTZ black hole has the  connection
\eqn\hswu{\eqalign{a_+ &= V^2_1 +{1\over 4\t^2}V^2_{-1}\cr
a_-&=0\cr}}
This is straightforward, as the $V^2_{\pm1}$ are  \slt\ elements. The BTZ holonomy can be encoded in the infinite set of traces
\eqn\bq{\Tr(\o_{BTZ}^n)~, ~~ n=2,3,\ldots}
where
\eqn\bqq{\o_{BTZ} = 2\pi\t\left(V^2_1 +{1\over 4\t^2}V^2_{-1}\right)}
All odd-$n$ traces vanish. The lowest even-$n$ traces are
\eqn\br{\eqalign{\Tr(\omega_{BTZ}^2) &= -8\pi^2\cr
\Tr(\omega_{BTZ}^4) &= {8\pi^4\over 5}(3\l^2-7)\cr
\Tr(\omega_{BTZ}^6) &= -{8\pi^6\over 7}(3\l^4-18\l^2+31)\cr}}

\subsec{The \mwil\ black hole}

Our  ansatz for a black hole with  spin-3 chemical potential is
\eqn\bs{\eqalign{a_+ &= V^2_{1} -{2\pi\ml\over k}V^2_{-1} -N(\l){\pi\mw\over 2k}V^3_{-2} + J\cr
a_-&= \mu N(\l)\left(a_+\star a_+ - {2\pi\ml\over 3k}(\l^2-1)\right) \cr}}
where
\eqn\btt{J = J_4V^4_{-3}+J_5V^5_{-4}+\ldots}
allows for an infinite series of higher-spin charges.  The solution is accompanied by the analogous  barred connection.  $N(\l)$ is a normalization factor,
\eqn\bttt{N(\l) = \sqrt{20\over(\l^2-4)}}
chosen to simplify comparison to the \slth\ results of \paps. In particular, truncating all spins $s>3$ gives a solution with the same generator normalizations and bilinear traces as the spin-3 black hole \akk\ of the \slth\ theory.\foot{To be clear, there is no pathology for $\l\leq 2$: we could easily rescale generators to eliminate any troublesome factors of $\l^2-4$.}

Suppressing the dependence on barred quantities, we think of this black hole as a saddle point contribution to the partition function
\eqn\buu{Z(\t,\a) = \Tr\left[e^{4\pi^2i(\t\ml+\a\mw)}\right]}
where we continue to define the potential as
\eqn\buuu{\a=\taub \mu}
where $\tau$ is the modular parameter of the boundary torus, defined via the identification $(z,\zb) \cong (z+2\pi \tau, \zb+2\pi \taub)$,
with $x^+=z,~ x^-=-\zb$.  This will once again be justified upon solving the holonomy equations, as the charges will satisfy the integrability condition
\eqn\bvv{{\p \Lc \over \p \alpha} = {\p \Wc \over \p \tau}}

The structure of this ansatz is understood as follows: $a_+$ is the asymptotically AdS$_3$ connection written in the ``highest weight gauge'' that was used to reveal the asymptotic \mwil\ symmetry in \refs{\HenneauxXG,\CampoleoniZQ}.
 The component $a_-$ is a traceless source term that deforms the UV asymptotics: by \bjjj,
\eqn\bu{a_- = \mu N(\l) V^3_2 + ({\rm subleading})}

Though similar in some ways, this \hsl\ black hole has some properties that are quite different from its \slth\ counterpart. First, there is an infinite set of holonomy constraints to solve, corresponding to enforcing smoothness across the horizon  of the metric and higher spin fields. Furthermore, solution of these constraints demands that {\it all} higher-spin charges are turned on. This is due to the structure of the $\Wc_\infty[\lambda]$ algebra. For instance, the $\mw\mw$ OPE has a term
\eqn\bx{\mw(z)\mw(0) \sim \ldots + {\mu J_4(0)\over z^2}+\ldots }
and likewise for OPEs of higher spin currents.

This sourcing of ever-higher spins is nicely on display in the bulk: we will find that without all spins turned on, \bvv\ is not satisfied by the solution of the holonomy equations.

\subsec{Holonomy}

For the \mwil\ black hole ansatz \bs, the holonomy matrix is
\eqn\cca{\o = 2\pi\Bigg[\tau a_+ - \a N(\l)\left(a_+\star a_+ - {2\pi\ml\over 3k}(\l^2-1)\right)\Bigg]}
with $a_+$ and $N(\l)$ as in \bs\ and \bttt, respectively. The holonomy constraints are
\eqn\ccb{\Tr(\o^n)=\Tr(\o_{BTZ}^n)~,\quad n= 2, 3, \ldots}

We proceed to solve \ccb\ perturbatively in $\a$. We assume a perturbative expansion of the form
\eqn\ccc{\eqalign{\ml &= \ml_0+\a^2\ml_2+\ldots\cr
\mw&= \a\mw_1+\a^3\mw_3+\ldots\cr
J_4 &= \a^2J_4^{(2)}+\a^4J_4^{(4)}+\ldots\cr
J_5 &= \a^3J_5^{(3)}+\a^5J_5^{(5)}+\ldots\cr}}
and so on for higher spins. The OPEs tell us that the spin-3 chemical potential $\mu$ sources the spin-4 current at $O(\mu^2)$, the spin-5 current at $O(\mu^3)$, and onwards, which this ansatz incorporates. Note the parity under sign flip of $\a$.

To exhibit the structure of the holonomy equations, and to set up our perturbative solution,
we write the terms that contribute at lowest perturbative order for each charge that appears in a given equation, ignoring all of the coefficients and displaying just the first four equations:

\eqn\ccd{\eqalign{\!\!\!\!\!\!n=2:~~ C_{BTZ}^{(2)}&= ~~\ml~+~\a\mw ~~+ \a^2J_4~~+\ldots\cr
\!\!\!\!\!\!n=3:~~ C_{BTZ}^{(3)}&= \a\ml^2+~~\mw ~~~+ ~\a J_4 ~~+ \a^2J_5 + \a^3J_6 +\ldots   \cr
\!\!\!\!\!\!n=4:~~C_{BTZ}^{(4)}&=   ~~\ml^2+\a\mw\ml ~+  ~~~J_4 ~~+ ~\a J_5 + \a^2J_6 + \a^3J_7+\a^4J_8  +\ldots \cr
\!\!\!\!\!\!n=5:~~ C_{BTZ}^{(5)}&=  \a\ml^3+~~\mw\ml ~+~\a J_4\ml + ~~~ J_5 +~ \a J_6 + \a^2J_7+\a^3J_8 + \a^4J_9+\a^5J_{10}+\ldots
}}
$C_{BTZ}^{(n)}$ stand for  the BTZ holonomies \br\ that are of course of order $\alpha^0$.
The ``$\ldots$'' denote terms that contribute at higher perturbative order (for instance, an $\a^2\ml^2$ term at $n=2$). At each value of $n$, two more charges enter at ever-higher orders in $\a$. In appendix B we write out the all-order holonomy equations up to $n=4$, with spins $J_5$ and higher set to zero for simplicity.

In the case that the gauge algebra is SL(N,R), as obtained by setting $\lambda=N$, the system of equations terminates at $n=N$.  For
the SL(3,R) theory studied in \paps\ this allowed the holonomy equations to be solved exactly as the solution of a cubic equation.  For general $\lambda$ we instead must proceed perturbatively.
In examining the structure of the equations \ccd\ one might be concerned by the fact
that at even(odd) orders in $\a$, {\it all} even(odd) $n$ equations contribute. This implies that the system is highly overconstrained, infinitely so, in fact; nevertheless it turns out that there is a consistent solution that satisfies the integrability condition, at least as far as we have checked.

We solve through $O(\a^8)$. Combining \ccc\ and \ccd, we see that we only need work up to $n=6$. The solution is
\eqn\fk{\eqalign{\!\!\!\!\!\!\ml &= -{k\over 8\pi\t^2} + {5k\over 6\pi\t^6}\a^2-{50k\over3\pi\t^{10}}~{\l^2-7\over\l^2-4}
\a^4+{2600k\over 27\pi\t^{14}}~{5\l^4-85\l^2+377\over (\l^2-4)^2}\a^6\cr
&\quad  -{68000k\over 81\pi\t^{18}}~{20\l^6-600\l^4+6387\l^2-23357\over(\l^2-4)^3}
\a^8+\ldots\cr
\!\!\!\!\!\!\mw &= -{k\over 3\pi\t^5}\a+{200k\over27\pi\t^{9}}~{\l^2-7\over\l^2-4}\a^3
-{400k\over 9\pi\t^{13}}~{5\l^4-85\l^2+377\over (\l^2-4)^2}\a^5\cr
&\quad+{32000k\over 81\pi\t^{17}}~{20\l^6-600\l^4+6387\l^2-23357\over(\l^2-4)^3}
\a^7+\ldots\cr
\!\!\!\!\!\!J_4 &= {35\over9\t^8}~{1\over\l^2-4}\a^2-{700\over9\t^{12}}~
{2\l^2-21\over(\l^2-4)^2} \a^4+{2800\over9\t^{16}}~{20\l^4-480\l^2+3189\over(\l^2-4)^3}
\a^6+\ldots\cr
\!\!\!\!\!\!J_5 &= {100\sqrt{5}\over9\t^{11}}~{1\over(\l^2-4)^{3/2}}\a^3 -{400\sqrt{5}\over 27\t^{15}}~{44\l^2-635\over(\l^2-4)^{5/2}}\a^5+\ldots\cr
\!\!\!\!\!\!J_6&={14300\over 81\t^{14}}~{1\over (\l^2-4)^2}\a^4+\ldots\cr    }}
When solving these equations the coefficients are obtained in a zigzag pattern:   first solve for the leading term in $\ml$, then that of $\mw$, then the subleading term in $\ml$, then the leading term in $J_4$, and so on.  From the solutions of $\Lc$ and $\Wc$ we readily confirm that the integrability equation \bvv\ is obeyed, although it has to be said that at our current level of understanding this appears as a minor miracle.  We take this to be powerful evidence that the holonomy prescription is the correct one for defining  higher spin black holes with consistent thermodynamics.\vs

To obtain the partition function we can  integrate either one of the equations
\eqn\fkka{   { \p \ln Z(\t,\a) \over \p \tau} = 4\pi^2 i \Lc~,\quad {\p \ln Z(\t,\a) \over \p \alpha} = 4\pi^2 i \Wc}
and thereby  arrive at the result quoted in the introduction:
\eqn\fkk{\eqalign{ \ln Z(\t,\a) & =
{i \pi k \over 2\tau} \Bigg[1 -{ 4\over 3}~{\alpha^2 \over \tau^4}
+{400 \over 27}~ { \lambda^2-7 \over \lambda^2-4}~{\alpha^4\over \tau^8} -{1600 \over 27}~{5\lambda^4 -85 \lambda^2 +377 \over (\lambda^2-4)^2}~ {\alpha^6 \over \tau^{12}}\cr
&\quad\quad\quad~   +{32000\over 81}~{20\l^6-600\l^4+6387\l^2-23357\over(\l^2-4)^3}~{\a^8\over\t^{16}}\Bigg]+ \ldots }}
%

This partition function is the main result of our bulk analysis. The black hole entropy $S$ can be obtained by applying standard thermodynamics:
\eqn\fkkb{ S = \ln Z(\t,\a) -4\pi^2i(\tau \Lc + \alpha \Wc -\taub \Lcb - \alphab \Wcb) }
Since the formula  for the entropy does not appear particularly illuminating we refrain from writing it.  Suffice it to say that at order $\alpha^0$ the
the entropy is $A/4G$, where $A$ is the area of the BTZ horizon, but at higher orders in $\alpha$ no geometric interpretation is evident.

\subsec{Comments and two checks}

Ignoring the factors of $\l^2-4$ for the moment --- which, we recall, can be normalized away --- the charges all take the form
\eqn\fnb{J_s=\t^{-s}\sum_{n=0}^{\infty}\left({\a\over\t^2}\right)^{2n+s-2}P_n^{(s)}(\l^2)~, ~~ s\geq3}
where we include $\mw \equiv J_3$. The  degree $n$ polynomials $P_n^{(s)}(\l^2)$ have zeroes that do not coincide with those of other values of $n$ or $s$, and so are unlikely to carry any significance.

A first check on our result \fk\ is the $\l$-independence of the first correction to $\mw$ and, by integrability, to $\ml$. This can be understood as following from the universal leading coefficient of the $\mw\mw$ OPE, which in our normalization is\foot{The sign convention adopted here differs from that in most CFT references, but turns out to be more convenient. }
\eqn\fl{\mw(z)\mw(0) \sim -{5k\over \pi^2}{1\over z^6}+\ldots}

In addition, we recall that the \hsl\ trace automatically mods out the effects of higher spin generators upon taking $\l=N$ for integer $N\geq 3$. So computing the holonomy in \hsl\ and then evaluating at $\l=N$ is identical to computing the holonomy in the SL(N,R) theory from the start.

To verify this, we have embedded the black hole with spin-3 chemical potential in the theories with Lie algebras $\cg=$SL(3,R), SL(4,R), SL(5,R). The results of these investigations match \fk\ exactly (modulo the non-existence of some of the $J$ charges in these cases). Furthermore, we have confirmed that the holonomy equations themselves reduce to those of SL(N,R) non-perturbatively (see appendix B for an example).

Another useful check is to consider hs[$\half$].   In this case,
the gauge algebra can be represented in terms of a Moyal product, with generators being even degree polynomials in two spinor variables \ProkushkinBQ.   All computations can then be carried out in this
framework, and the results precisely agree with \fk\ at $\lambda = 1/2$.

\newsec{ $\l=0,1$:  comparison to CFT}

Recall that \mwil\ is the asymptotic symmetry of the AdS$_3$ vacuum of the \hsl\ theory, and possesses \hsl\ as its wedge subalgebra in the $c\rar\infty$ limit, as discussed in \GaberdielWB. In anticipation of an application of our results to the holographic realm, we switch gears and study two CFT realizations of \mwil\ symmetry. One is a theory of free bosons at $\l=1$, and the other, a theory of free fermions at $\l=0$. We compute their exact partition functions in the presence of a spin-3 chemical potential; the perturbative expansions match \fkk.
We defer a discussion of why this should be, and its intriguing implications, to the next section.

For easy reference, the gravity results for the leftmoving part of the partition function are
\eqn\mre{\eqalign{\!\!\!\!\l=1:\quad&  \ln Z(\tau,\alpha) = {i \pi k \over 2\tau} -{2 i  \pi k \over 3}{\alpha^2 \over\tau^5} +{400 i  \pi k \over 27}{\alpha^4 \over \tau^9} -{8800 i \pi k \over 9} {\alpha^6 \over \tau^{13} }+{10400000i \pi k\over81}{\a^8\over\t^{17}} +\cdots   \cr
\!\!\!\!\l=0:\quad& \ln Z(\tau,\alpha) = {i\pi k \over 2\tau}- {2i\pi k\over 3}{\alpha^2 \over \tau^5}+{350 i \pi k \over 27}{\alpha^4 \over \tau^9}-{18850 i \pi k \over 27}{\alpha^6 \over \tau^{13}} +{5839250i \pi k\over81}{\a^8\over\t^{17}}+\cdots}}
We ignore the rightmoving part henceforth and focus on a single chiral sector.

\subsec{$\l=1$: Free bosons}

A theory of $D$ free complex bosons has global \mwio\ symmetry with central charge $c=2D$ \refs{\BakasRY,\HullSA,\HullKF}.  The symmetry algebra is also known \prs\ as $\mw_{\infty}^{\rm PRS}$, and can be written in a linear basis, unlike the \mwil\ algebra for other values of $\l$.

The complex bosons have OPE
\eqn\je{ \p \phib^i(z_1 ) \p \phi_j(z_2 )  \sim  - {\delta^i_j \over (z_1-z_2)^2}~,\quad i, j = 1, 2, \ldots D }
The stress tensor and spin-3 current are (summations implied)
\eqn\jf{\eqalign{T &= - \p \phib^i \p \phi_i \cr
\Wc  &= ia \left(\p^2 \phib^i \p \phi_i - \p \phib^i \p^2 \phi_i\right)\cr}}
where  $a$ is some normalization constant. $\Wc$ is Virasoro primary, as it should be to match with the spin-3 current of our bulk theory. The other higher spin currents are also quadratic in the scalars but include more derivatives; the linearity of the symmetry algebra follows from the quadratic nature of these currents.

To fix $a$, we compute the leading part of the $\Wc\Wc$ OPE:
\eqn\ji{ \Wc(z)\Wc(0)\sim -{4 a^2 D \over z^6} +\cdots   }
Since our standard normalization is
\eqn\jj{  \Wc(z)\Wc(0) \sim -{5k \over \pi^2 z^6}}
to match up we take
\eqn\jk{ a = \sqrt{ 5k \over 4\pi^2 D} }
Trading $D$ for $k$ according to  $c=6k = 2D$, we  have
\eqn\jl{   a =  \sqrt{5 \over 12\pi^2 } }

Now  we write down mode expansions. We work
on the cylinder with coordinate  $w= \sigma_1 +i \sigma_2$, related
to the plane coordinate $z$ as $ z=e^{-iw}$. We  suppress the $i$ indices for the time being. We write
\eqn\jcf{ \p \phi(w) = - \sum_m \beta_m e^{imw}~,\quad  \p \phib(w) = - \sum_m \betab_m e^{imw}}
where the modes obey
\eqn\jch{ [\betab_m,\beta_n]= m \delta_{m,-n}= [\beta_m,\betab_n]}

The normal ordered stress tensor is
\eqn\jci{\eqalign{ T&= -\sum_{m=1}^\infty \Big(\betab_{-m} \beta_m +\beta_{-m} \betab_m\Big) + {k\over 4} +{\rm nonconstant} }}
Since what appears in the partition function is the zero mode of the stress tensor, only the constant terms are relevant here. In the remainder of this section we drop the ground state $k/4$ term, as it plays no role in the high temperature expansion that we are interested in.

As in \papa, the  quantity $\Lc$ is related to the constant part of the stress tensor as
\eqn\jcj{ \Lc = -{1\over 2\pi} T}
and so we   have
\eqn\jck{ \Lc = {1\over 2\pi} \sum_{m=1}^\infty \Big(\betab_{-m} \beta_m +\beta_{-m} \betab_m\Big) }

Now consider the mode expansion of the spin-3 charge, which is the constant part of $\Wc$.  We get, after normal ordering,
\eqn\jcl{ \Wc =2a \sum_{m=1}^\infty  m\Big( \betab_{-m}\beta_m -\beta_{-m} \betab_m\Big)}

We can think of the states as being described by arbitrary numbers of
positively and negatively charged particles.  For example,  a state of the form
\eqn\jt{   |n_m, \overline{n}_m\rangle=  (\beta_{-m})^{n_m} (\betab_{-m})^{\overline{n}_m}|0\rangle  }
obeys
\eqn\ju{\eqalign{  \Lc  |n_m, \overline{n}_m\rangle& = {1\over 2\pi} m(\overline{n}_m+n_m) |n_m, \overline{n}_m\rangle \cr
 \Wc  |n_m, \overline{n}_m\rangle & = 2a m^2(\overline{n}_m-n_m)  |n_m, \overline{n}_m\rangle} }

It is now elementary to compute the partition function
\eqn\jua{ Z(\tau,\alpha) = \Tr \Big[ e^{4\pi^2 i (\tau \Lc + \alpha \Wc) } \Big]}
and we obtain
\eqn\jw{\ln Z(\tau,\alpha) = - D \sum_{m=1}^\infty \left[\ln \left(1 - e^{ 2\pi i \tau m -8\pi^2 ia \alpha m^2  } \right) + \ln \left(1 - e^{ 2\pi i \tau m +8\pi^2 ia \alpha m^2   } \right)  \right] }

In the high temperature regime, $\tau_2 \rt 0$,  we can convert the sum to an integral.
This gives
\eqn\jy{ \ln Z(\tau,\alpha) = -{3 ik \over 2\pi \tau} \int_0^\infty\! dx \left[ \ln \left(1-e^{-x + {2 ia \alpha  \over \tau^2}x^2 }\right) +
\ln \left(1-e^{-x - {2 ia \alpha  \over \tau^2}x^2 }\right)\right]}
where we also used $D=3k$.

It is straightforward to expand in powers of $\alpha$ and do the integrals:
\eqn\jz{ \ln Z(\tau,\alpha) = {i \pi k \over 2\tau} -{2 i \pi k  \over 3}{\alpha^2 \over\tau^5} +{400 i\pi k  \over 27}{\alpha^4 \over \tau^9} -{8800 i\pi k  \over 9} {\alpha^6 \over \tau^{13} } +{10400000i\pi k\over81}{\a^8\over\t^{17}}+\cdots    }
This  agrees precisely with the gravity result \mre.

\subsec{$\l=0$: Free fermions}
A theory of $D$ free complex fermions furnishes $\mw_{1+\infty}$ symmetry  with central charge $c=D$ \BergshoeffYD. The $\mw_{1+\infty}$ algebra has spins $s=1,2,3,\ldots$, and is related to $\mw_{\infty}[0]$ by a constraint that eliminates the spin-1 current. In the following we will proceed by using a chemical potential to demand that the spin-1 charge is set to zero in the partition function.

The fermions have OPE
\eqn\ma{ \psib^i(z_1) \psi_j(z_2) \sim {\delta^i_j\over z_1 -z_2} }
The stress tensor is
\eqn\mb{ T = -{1\over 2} \psib^i \p \psi_i -{1\over 2} \psi_i \p \psib^i}

According to \BergshoeffYD\ the relevant  spin-3 current is, up to a normalization constant $b$ that we will fix in a moment,
\eqn\mc{ \Wc = i b \big( \p^2 \psib^i \psi_i -4 \p \psib^i \p \psi_i + \psib^i \p^2 \psi_i \big)}
As noted in \PopeKC, this current is not primary, as the
$T\Wc$ OPE contains an extra term of the form $J/z^4$ where $J$ is the spin-1 current.   From this point of view it is clear that to compare with the bulk we need to set $J=0$ so that $\Wc$ will appear
as effectively primary.

Now we consider the leading part of the $\Wc \Wc$ OPE, which is
\eqn\me{ \Wc(z) \Wc(0) \sim -{24  D b^2 \over z^6} +\cdots   }
where we've now taken $D$ complex fermions.
To match to our usual normalization we should take
\eqn\mg{ b = {\sqrt{5 \over 144\pi^2 }} }
where we've used $D=c=6k $.

The mode expansion on the cylinder is\foot{We suppress the $i$ indices.}
\eqn\mj{ \psib(w) = e^{-{i\pi \over 4}} \sum_m \barb_m e^{imw}~,\quad
 \psi(w) = e^{-{i\pi \over 4}} \sum_m b_m e^{imw} }
with
\eqn\mja{ \{ \barb_m, b_n\} = \delta_{m,-n}}
Ignoring the zero point energy and nonconstant terms, the normal ordered stress tensor is
\eqn\mk{ T(w) =  -\sum_{m=1}^\infty m   \big( \barb_{-m}b_m + b_{-m}\barb_m\big)    }
so
\eqn\mka{ \Lc = {1\over 2\pi} \sum_{m=1}^\infty m   \big( \barb_{-m}b_m + b_{-m}\barb_m\big)    }
Similarly, the  spin-3 current is
\eqn\mll{ \Wc = -6b \sum_{m=1}^\infty m^2 \big(\barb_{-m} b_m - b_{-m}\barb_{m}\big)  }

 A state of the form
\eqn\mla{   |n_m, \overline{n}_m\rangle=  (b_{-m})^{n_m} (\barb_{-m})^{\overline{n}_m}|0\rangle  }
obeys
\eqn\mlb{\eqalign{  \Lc  |n_m, \overline{n}_m\rangle& = {1\over 2\pi} m(\overline{n}_m+n_m) |n_m, \overline{n}_m\rangle \cr
 \Wc  |n_m, \overline{n}_m\rangle & = -6b m^2(\overline{n}_m-n_m)  |n_m, \overline{n}_m\rangle} }

Finally, we need to consider the spin-1 charge operator, which is
$Q \sim  \int \psib \psi$.
Our precise definition of  the charge operator is
\eqn\maa{ Q|n_m, \overline{n}_m\rangle = (\barn_m - n_m)|n_m, \overline{n}_m\rangle}

In the thermodynamic (high temperature) limit it won't matter whether we impose $Q=0$ as an exact condition on states or as an expectation value;  the latter is more convenient since it can be imposed by including a chemical potential for $Q$ and tuning it appropriately.
The partition function including a chemical potential for $Q$ is
\eqn\mna{\eqalign{ Z(\tau,\alpha,\gamma) & = \Tr ~ \Big[e^{4\pi^2 i [ \tau \Lc + \alpha \Wc ]+i \gamma Q }\Big]}}
We calculate this to be
\eqn\mo{\ln Z(\tau,\alpha,\gamma) =  D\sum_{m=1}^\infty \left[\ln \left(1 + e^{( 2\pi i \tau m -24\pi^2 ib  \alpha m^2+i\gamma  ) } \right) + \ln \left(1 + e^{( 2\pi i \tau m +24 \pi^2 ib  \alpha m^2-i\gamma  ) } \right)  \right] }
Converting the sum to an integral  we have
\eqn\mr{ \ln Z(\tau,\alpha,\gamma) = {3 ik \over \pi \tau} \int_0^\infty\! dx \left[ \ln \left(1+e^{-x + {6ib   \alpha  \over \tau^2}x^2 +i\gamma }\right) +
\ln \left(1+e^{-x - {6 ib \alpha  \over \tau^2}x^2 -i\gamma}\right)\right]}

We now fix $\gamma$ by demanding charge neutrality.  The charge is obtained by differentiating with respect to $\gamma$ and so we need
\eqn\mra{ 0 = \int_0^\infty dx\left[{1 \over e^{-x-i\eps x^2 -i\gamma}+1} -  {1 \over e^{-x+i\eps x^2 +i\gamma}+1}   \right]}
where we defined
\eqn\mrb{ \eps = {6b\alpha\over \tau^2}  }
Solving perturbatively gives
\eqn\mrc{ \gamma = -{\pi^2 \over 3} \eps +{16 \pi^4 \over 9}\eps^3 -{448 \pi^6 \over 9}\eps^5 +{1254656\pi^8\over405}\eps^7+\cdots}

We now plug this into \mr, expand in $\eps$, and compute the integrals.  After inserting the value of $b$ given above, we find
\eqn\mrd{ \ln Z(\tau,\alpha) = {i\pi k \over 2\tau}- {2i\pi k\over 3}{\alpha^2 \over \tau^5}+{350 i \pi k \over 27}{\alpha^4 \over \tau^9}-{18850 i \pi k \over 27}{\alpha^6 \over \tau^{13}}+{5839250i\pi k\over81}{\a^8\over\t^{17}} +\cdots }
This  agrees precisely with the gravity result \mre.

\newsec{Implications for higher spin AdS$_3$/CFT$_2$ duality}

We now consider what lessons can be drawn from the agreement between our bulk gravity computations and those for free bosons and fermions. For this discussion, let us make the assumption that the agreement will persist to all order in $\alpha$.

Symmetry obviously plays a powerful role in determining these partition functions.  The most likely explanation for why we see agreement is that the answer is fixed by symmetry.   On the bulk side, our black hole solutions just involve the non-propagating bulk fields described by the Chern-Simons action, and not the additional scalar fields that arise in the context of the conjecture \GaberdielPZ.  Since the topological sector is what gives rise to the asymptotic symmetry algebra, it seems plausible that the physical properties of solutions that lie in this sector are fixed by symmetry.

On the CFT side, a symmetry argument would probably proceed along the following lines.\foot{We thank Matthias Gaberdiel for discussions of these matters.} The partition functions we compute are determined, at order $\alpha^n$, by the n-point correlation functions of spin-3 currents on the torus.  We are interested in the high temperature behavior of these 
correlation functions.   Performing modular transformations term by term,
the leading high temperature behavior will be related to correlation functions at low temperature, which are evaluated on the infinite cylinder, or equivalently the plane.   Finally, the correlation functions of spin-3 currents on the plane can be computed from the OPEs.   Thus, given the OPEs, we expect that we should be able to compute the partition function in the high temperature limit, and it should agree with our gravity result. 

Nothing in this argument depends on considering the special cases $\lambda =0,1$ where we could compute explicitly in terms of free fermions/bosons, 
and so for any $\lambda$ we expect agreement between our black hole partition function and that for the coset CFT \inb\ in the high temperature regime.   From this point of view, it would be especially interesting to try to understand and match the subleading asymptotics.

As was already mentioned in the introduction, our final  result should be thought of as a Cardy formula for CFTs with $\Wc_\infty[\lambda]$ symmetry and with large central charge.

Although we have argued that our successful matching of partition functions
in terms of free fermions and bosons is a consequence of symmetry, it is interesting to note that for $\lambda=0$ it is believed that the theory
\inb\  is in fact fully equivalent to free fermions with a singlet constraint.$^8$   It would therefore be very interesting to carry out further
tests of the AdS/CFT duality \GaberdielPZ\ at $\lambda =0$. 

We conclude with one final simple observation.   Both for $\lambda =0$ and $\lambda =1$, in the free fermion/boson theories, there are natural candidates for operators dual to the scalar fields that appear on the bulk side of the duality \GaberdielPZ.  These bulk scalars are dual to spinless operators in the CFT of dimension
$\Delta = 1\pm \lambda$.  At $\lambda =0$ we have the free fermion
operator $\psi(z)\tilde{\psi}(\zb)$, and at $\lambda =1$ we have
the free boson operator $\p \phi(z) \overline{\p} \phi(\zb)$, both of which have the appropriate dimension.

\vskip .3in

\noindent
{ \bf Acknowledgments}

\vskip .3cm

 This work was supported in part by NSF grant PHY-07-57702.
 We thank Martin Ammon, Alejandra Castro,  Matthias Gaberdiel, Michael Gutperle, Finn Larsen, and  Alex Maloney for discussions.  P.K. thanks the Aspen Center for Physics for hospitality during the completion of this work.

\appendix{A}{The \hsl\ Lie algebra}
The \hsl\ structure constants are given as
\eqn\bc{g_u^{st}(m,n;\lambda) = {q^{u-2}\over2(u-1)!}\phi_u^{st}(\lambda)N_u^{st}(m,n)}
where
\eqn\bd{\eqalign{N_u^{st}(m,n) &= \sum_{k=0}^{u-1}(-1)^k
\left(\matrix{u-1 \cr k}\right)
[s-1+m]_{u-1-k}[s-1-m]_k[t-1+n]_k[t-1-n]_{u-1-k}\cr
\phi_u^{st}(\lambda) &= \ _4F_3\left[\matrix{\half + \lambda ~,~   \half - \lambda  ~,~ {2-u\over 2} ~ ,~ {1-u\over 2}\cr
{3\over 2}-s ~ , ~~ {3\over 2} -t~ ,~~ \half + s+t-u}\Bigg|1\right]\cr}}
We make use of the descending Pochhammer symbol,

\eqn\hswee{[a]_n  = a(a-1)...(a-n+1)}
$q$ is a normalization constant that can be scaled away by taking $V^s_m \rar q^{s-2}V^s_m$. As in much of the existing literature, we choose to set $q=1/4$.

We note a handful of useful properties of the structure constants:
\eqn\hswsw{\eqalign{&\phi^{st}_u\left(\half\right)=\phi^{st}_2(\l)=1\cr
&N^{st}_u(m,n)=(-1)^{u+1}N^{ts}_u(n,m)\cr
&N^{st}_u(0,0)=0\cr
&N^{st}_u(n,-n)=N^{ts}_u(n,-n)\cr}}
The first three of these imply, among other things, the isomorphism hs[$\half$]~$\cong$~hs(1,1); that the lone star product can be used to define the \hsl\ Lie algebra; and that all zero modes commute.

\appendix{B}{Holonomy equations with $J_4\neq 0$}
Here  we present the holonomy equations
\eqn\aca{\Tr(\o^n)=\Tr(\o_{BTZ}^n)}
for the black hole connection \bs, up to $n=4$.  In the interest of clarity and space, we only include charges up to $J_4$. We find
\eqn\wmi{\eqalign{n=2:~ 0 &= \a^2J_4\big(144k^2(\l^2-9)\big) -1792\pi^2\alpha^2\ml^2- 504\pi\alpha\tau k\mw \cr&-168\pi\tau^2k\ml -21k^2\cr
n=3:~ 0 &= \a J_4\Big(36k^2(\l^2-4)(\l^2-9)\big[80\pi\a^2\ml(\l^2-16)+9k\t^2(\l^2-4)\big]\Big)\cr
&-40\a^3\pi^2\Big[45\mw^2k(5\l^4-65\l^2+264)+256\ml^3\pi(\l^2-4)(\l^2-16)\Big]\cr
&-4320\a^2\mw\ml\pi^2\tau k(\l^2-4)(4\l^2-29)\cr
&-4032\a\ml^2\pi^2\tau^2k(\l^2-4)^2\cr
&-189\mw\pi\tau^3k^2(\l^2-4)^2\cr
n=4:~ 0 &=\a^4J_4^2\Big(57600k^4(\l^2-4)(\l^2-9)\big[35\l^6-1330\l^4+21707\l^2-134748\big]\Big)\cr
&-J_4\Big(624k^2(\l^2-4)(\l^2-9)\Big[12800\a^4\ml^2\pi^2(7\l^4-199\l^2+1788)\cr
&+8400\a^3\mw\pi\tau k(5\l^4-95\l^2+636)\cr
&+23760\a^2\ml\pi\tau^2 k(\l^2-4)(\l^2-11)+99\tau^4k^2(\l^2-4)^2\Big]\Big)\cr
&+665600\a^4\ml\pi^3\Big[75\mw^2k(\l^2-9)(5\l^4-95\l^2+636)\cr&+352\ml^3\pi(\l^2-4)(\l^4-17\l^2+100)\Big]\cr
&+131788800\a^3\mw\ml^2\pi^3\tau k(\l^2-4)\Big[3\l^4-51\l^2+244\Big]\cr
&+137280\a^2\pi^2\tau^2 k(\l^2-4)\Big[64\ml^3\pi(\l^2-4)(11\l^2-71)\cr&+45\mw^2k(5\l^4-65\l^2+264)\Big]\cr
&+823680\a\mw\ml\pi^2\tau^3k^2(\l^2-4)^2\Big[23\l^2-123\Big]\cr
&-9009k^2(3\l^2-7)(\l^2-4)^2(k+8\pi\ml\tau^2)(k-8\pi\ml\tau^2)\cr}}

We have organized these equations so as to reveal the $J_4$ dependence. Comparison to \ccd\ reveals the underlying structure discussed in the main text.

It is instructive to ask what happens when we take $\l=3$. Since this reduces to the SL(3,R) case, in which there are only two independent holonomy equations, one requires that the $n=4$ equation should vanish on account of the other two, and that any $J_4$ dependence should drop out of these equations.

The latter is evident upon inspection. And indeed, taking $\l=3$ reduces the mess of $n=4$ to be proportional to the $n=2$ equation by a finite factor. Both the $n=2$ and $n=3$ equations reduce to those in \papa\ (see e.g. equation (5.14) there).

\listrefs
\end